\documentclass{vgtc}                          

\graphicspath{{figures/}{pictures/}{figs/}{images/}{./}} 

\usepackage{times}                     

\usepackage{tabu}                      
\usepackage{booktabs}                  
\usepackage{lipsum}                    
\usepackage{mwe}                       
\usepackage{multirow}
\usepackage{tikz}
\usetikzlibrary{fit,backgrounds}

\usepackage{mathptmx}                  
\usepackage{caption}
\usepackage{subcaption}
\usepackage{cite}
\PassOptionsToPackage{hyphens}{url}

\newcommand{\tikzmarknode}[2]{%
  \tikz[remember picture,baseline] \node[anchor=base] (#1) {#2};%
}

\definecolor{deepblue}{HTML}{175858}
\definecolor{skyblue}{HTML}{289C9C}
\definecolor{iceblue}{HTML}{31BEBE}
\definecolor{NavyBlue}{HTML}{000080}

\newcommand{\expv}[1]{\textcolor{iceblue}{\textbf{#1}}}

\onlineid{1143}

\vgtccategory{Research}

\vgtcinsertpkg

\title{Scope Meets Screen: Lessons Learned in Designing Composite Visualizations for Marksmanship Training Across Skill Levels}

\author{Emin Zerman\thanks{e-mail: emin.zerman@miun.se} %
\and Jonas Carlsson\thanks{e-mail:j\_c\_kc@live.se} %
\and M{\aa}rten Sj\"ostr\"om\thanks{e-mail:marten.sjostrom@miun.se}}
\affiliation{\scriptsize Department of Computer and Electrical Engineering \\ Mid Sweden University, Sundsvall, Sweden}

\teaser{
  \centering
  \includegraphics[width=0.95\linewidth]{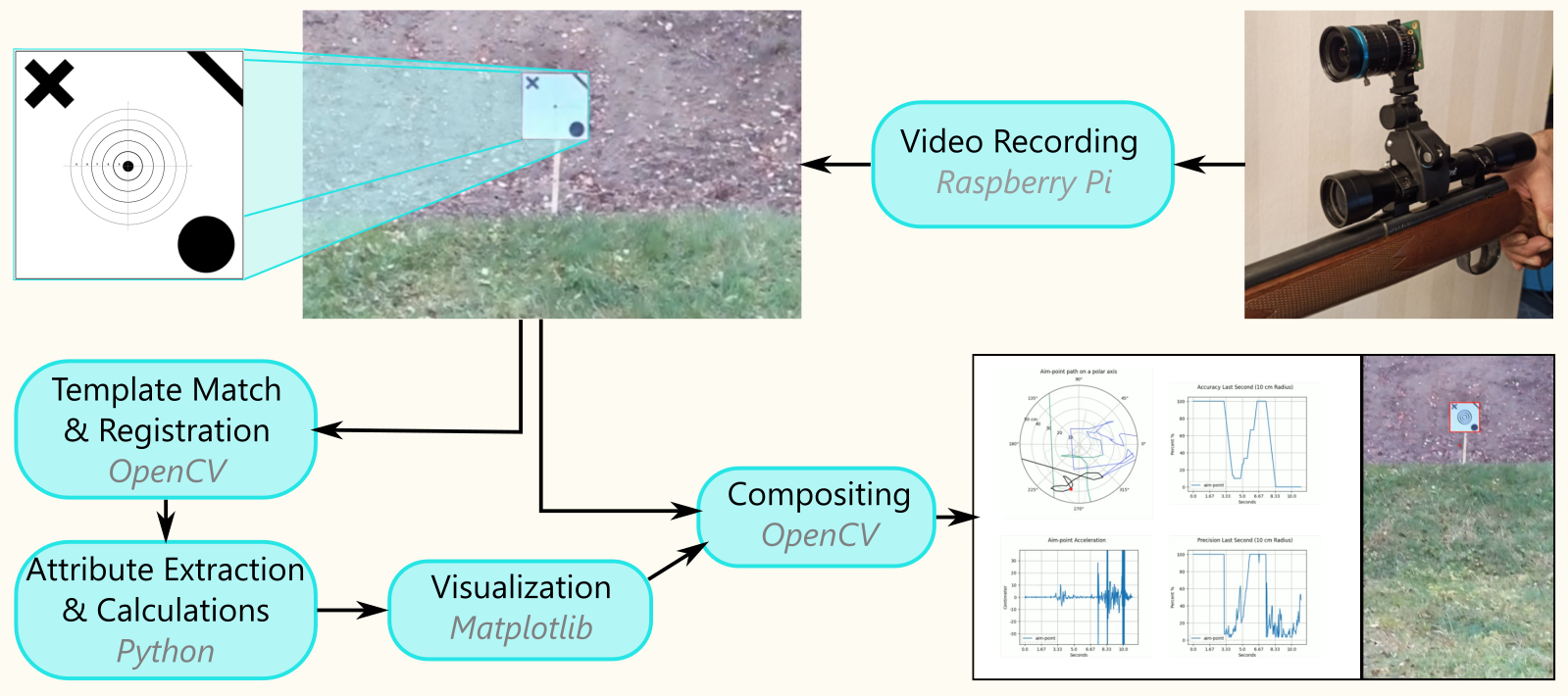}
  \caption{Overview of the developed system for creating composite visualizations.}
  \label{fig:teaser}
}

\abstract{
  Marksmanship practices are required in various professions, including police, military personnel, hunters, as well as sports shooters, such as Olympic shooting, biathlon, and modern pentathlon. The current form of training and coaching is mostly based on repetition, where the coach does not see through the eyes of the shooter, and analysis is limited to stance and accuracy post-session. In this study, we present a shooting visualization system and evaluate its perceived effectiveness for both novice and expert shooters. To achieve this, five composite visualizations were developed using first-person shooting video recordings enriched with overlaid metrics and graphical summaries. These views were evaluated with 10 participants (5 expert marksmen, 5 novices) through a mixed-methods study including shot-count and aiming interpretation tasks, pairwise preference comparisons, and semi-structured interviews. The results show that a dashboard-style composite view, combining raw video with a polar plot and selected graphs, was preferred in 9 of 10 cases and supported understanding across skill levels. The insights gained from this design study point to the broader value of integrating first-person video with visual analytics for coaching, and we suggest directions for applying this approach to other precision-based sports.
} 


\begin{document}


\firstsection{Introduction}

\maketitle

Sports training seeks rapid and situated feedback~\cite{handford1997skill, baca2006rapid, sigrist2013augmented} where the coach requires understanding of the situation and provides feedback to improve the trainees' performance. Visualizations can satisfy this important need for tracking and visual feedback~\cite{hoferlin2010video, kosmalla2017climbvis, perin2018state, fu2023hoopinsight}, as learning complex motor tasks were found to be positively affected by concurrent or rapid terminal visual feedback~\cite{sigrist2013augmented}.

Target shooting is no different from other sports in terms of requirements, albeit the most commonplace practice is to supervise the trainee's stance, grip, and breathing during shooting~\cite{knight2007perfect, brown2018stance} and analyze the accuracy by inspecting bullet holes after shooting~\cite{johnson2008crucial}. An augmented first-person view can increase the efficiency of the training process where additional information with respect to trainee's aim are visible along with the first-person video recording. Video visualizations require extracting meaningful information from the video through recognition, detection, and tracking to provide insight~\cite{borgo2012state, hoferlin2010video}. Scientific studies for visualizations on target shooting mostly focuses on the realization of the engineering tasks, such as computer vision~\cite{haug2018computer}, extraction of relevant features~\cite{ball2003body, jansson2019predicting}, noise removal from extracted features~\cite{koshiba2020video}, and video streaming~\cite{goransson2021exploring}. Existing training systems either rely on laser-based dry-fire (without ammo) applications~\cite{SCATT, ball2003body}, simulators~\cite{jansson2019predicting}, and virtual reality (VR)~\cite{munoz2020psychophysiological}. Dry-fire and VR systems lack the physical recoil, reducing training realism. 
Moreover, they do not provide the video footage, 
omitting important visual cues for the instructor and rapid terminal feedback for the trainee.

\begin{figure*}[tbp]
  \centering
  \begin{subfigure}[b]{0.32\linewidth}
  	\centering
  	\includegraphics[width=\textwidth, alt={The visualization of a raw video footage of a target in the field}]{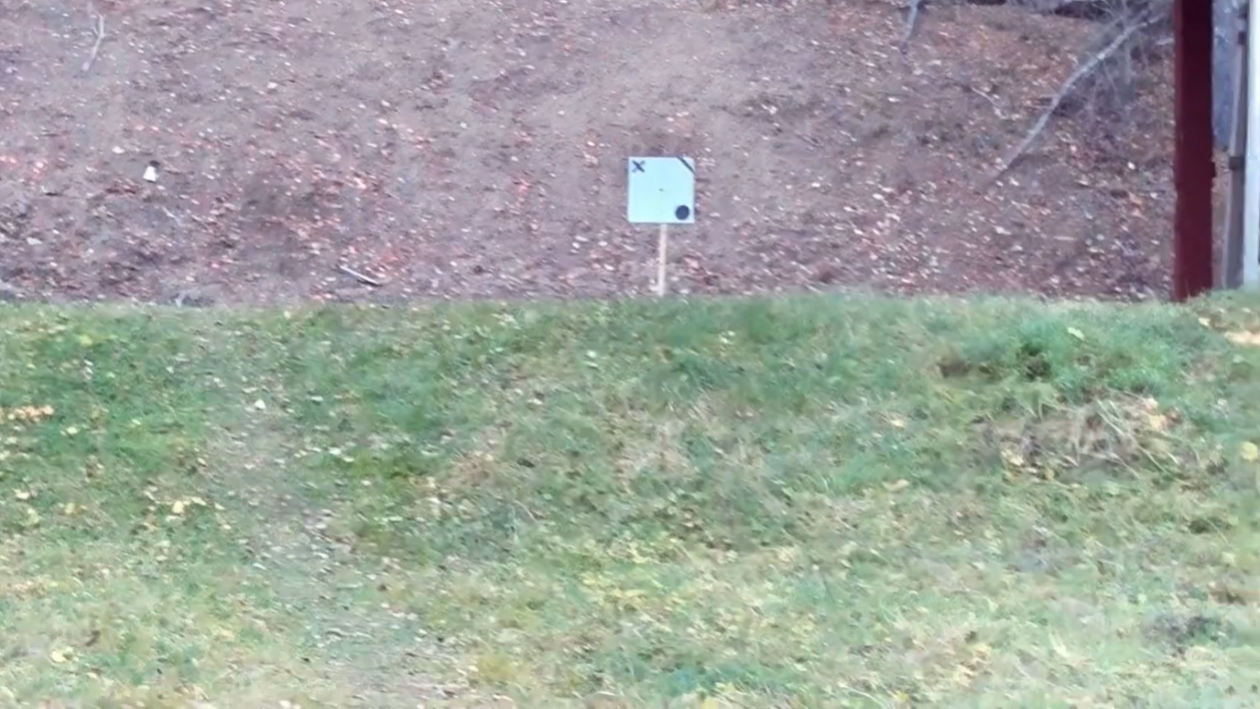}
  	\caption{Vis \#1: The raw video footage.}
  	\label{fig:design_subfigs_a}
  \end{subfigure}%
  \hfill%
  \begin{subfigure}[b]{0.32\linewidth}
  	\centering
  	\includegraphics[width=\textwidth, alt={A compound visualization where the original video is shown in the center, surrounded by a solid color and some metrics related to accuracy and precision shown on the left as text.}]{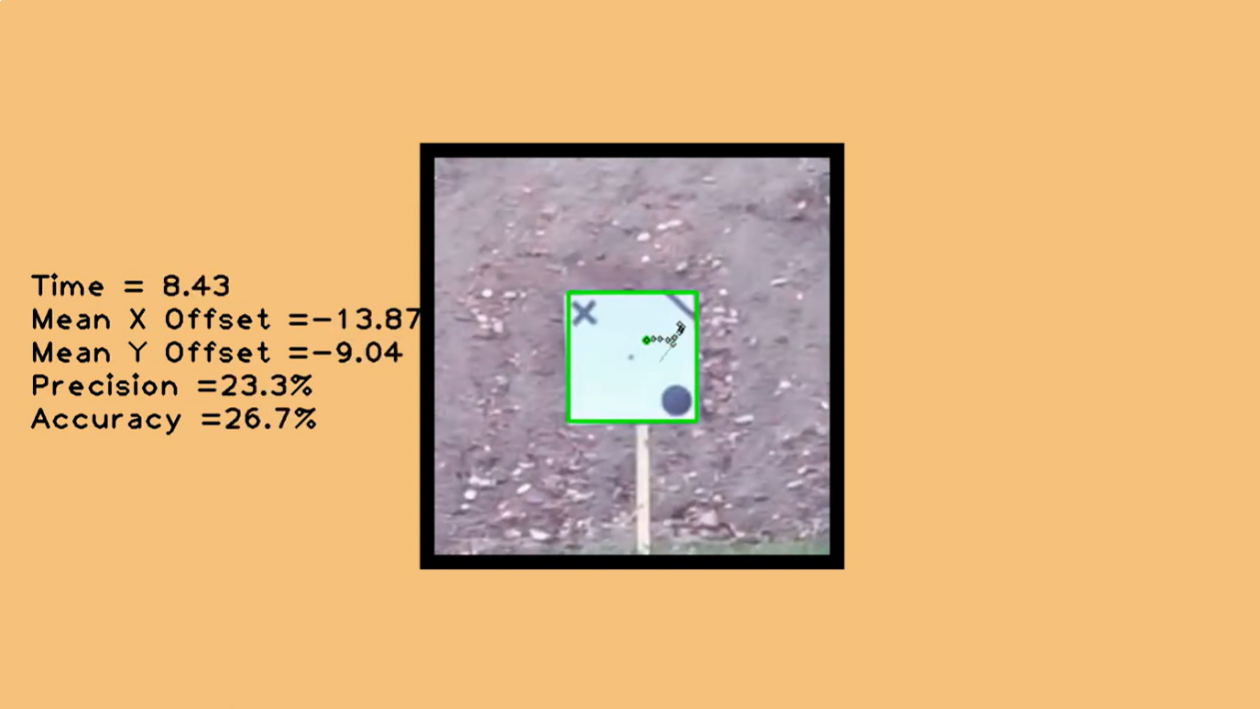}
  	\caption{Vis \#2: Stabilized video with text overlay.}
  	\label{fig:design_subfigs_b}
  \end{subfigure}%
  \hfill%
  \begin{subfigure}[b]{0.32\linewidth}
  	\centering
  	\includegraphics[width=\textwidth, alt={Time-series visualization of selected factors, including postion, distance of the aim from target, accuracy, and precision.}]{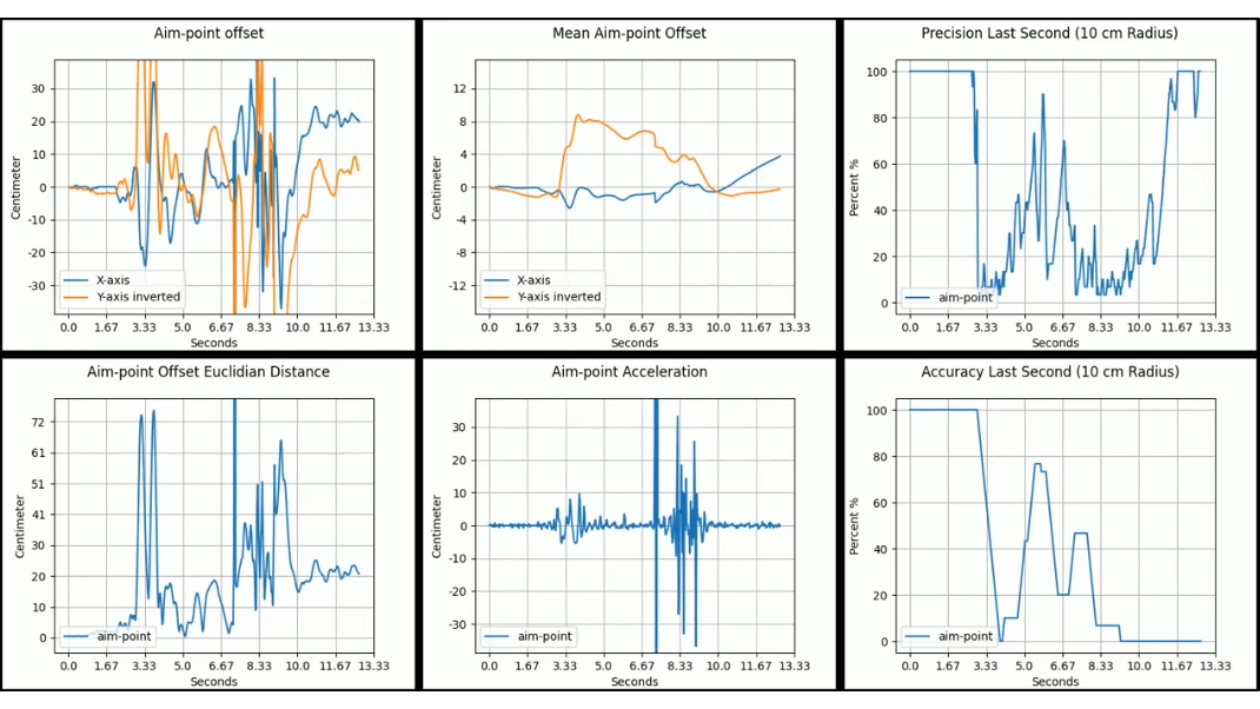}
  	\caption{Vis \#3: Time-series visualization of selected attributes.}
  	\label{fig:design_subfigs_c}
  \end{subfigure}%
  \\%
  \hfill%
  \begin{subfigure}[b]{0.32\linewidth}
  	\centering
  	\includegraphics[width=\textwidth, alt={A polar plot showing the aim point and the target, where the historical aim point is also shown as a ``tail''.}]{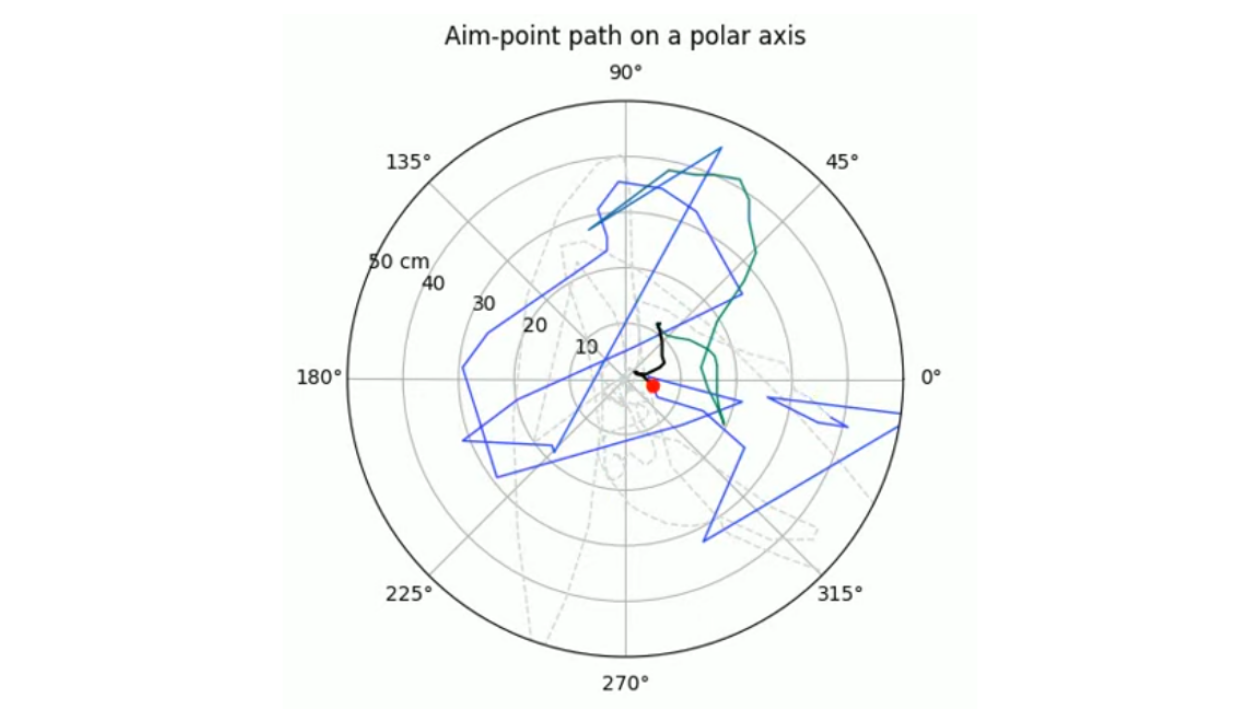}
  	\caption{Vis \#4: A polar plot with the aim point and the target.}
  	\label{fig:design_subfigs_d}
  \end{subfigure}%
  \hfill%
  \begin{subfigure}[b]{0.32\linewidth}
  	\centering
  	\includegraphics[width=\textwidth, alt={A dashboard style composite visualization, where the polar plot, accuracy, and precision are presented next to the raw video.}]{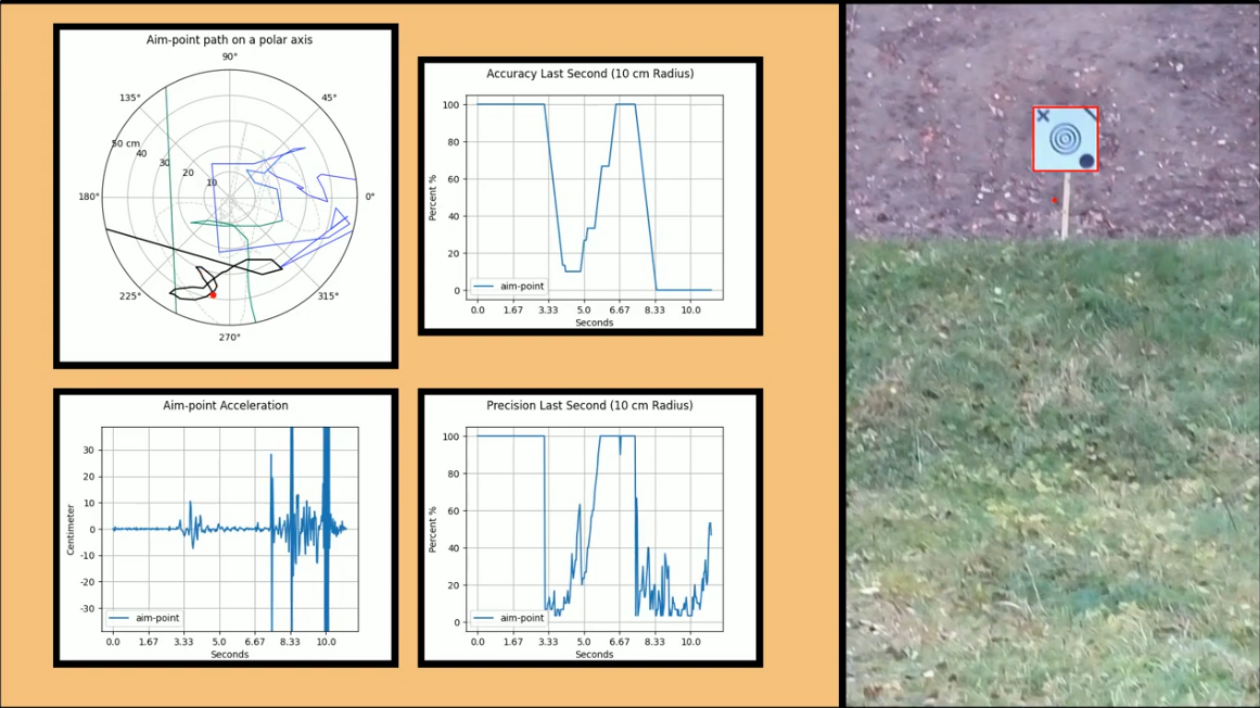}
  	\caption{Vis \#5: Combined composite visualization}
  	\label{fig:design_subfigs_e}
  \end{subfigure}%
  \hfill{~~}%
  \caption{Screenshots of the five visualization alternatives considered (cf. \cref{subsec:compositeVis} for explanations), including raw video and composite visualizations which include a combination of information visualization and the raw video footage. See supplementary materials for videos.
  }
  \label{fig:design_subfigs}
\end{figure*}

Composite visualizations~\cite{javed2012exploring} help provide an overview of the status through a logical collection of various visualizations and visual representations within the same view. The most commonly used composite visualization techniques in video-based visual analytics and sports training are juxtaposition (i.e., side-by-side placement of different visualizations) and superimposition (i.e., augmentation or overlaying of additional information on a ``base'' view)~\cite{hoferlin2010video, kosmalla2017climbvis, perin2018state, fu2023hoopinsight, jansson2019predicting, lin2023ball, lee2024sportify}. These techniques have become incredibly popular for sports media events to explain tactics or provide statistics to spectators without taking viewers' attention from sports games~\cite{lo2022stats, fifa_semi_2022, perry2003all, lin2023ball}, 
when spatial registration and timeliness are handled properly~\cite{lo2023sports}. Nevertheless, using composite visualizations 
for marksmanship training have rarely been addressed. 

This paper, based on and expanding Carlsson’s MSc thesis~\cite{carlsson2023shooting}, demonstrates  the utility of composite video visualizations in marksmanship training and distills design lessons relevant to broader sports training contexts through the analysis of qualitative user feedback.
For completeness, we introduce composite video visualizations 
with scope footage 
(\cref{sec:designing}), 
report the findings from a mixed-methods user study with expert and non-expert shooters (\cref{sec:evaluation}), and discuss implications for visualization design and coaching applications in related sports domains (\cref{sec:discussion}).

\section{Designing Composite Visualizations for Aiming} \label{sec:designing}
We designed and implemented a complete system to create composite visualizations for aiming. The 
steps included mounting a Raspberry Pi camera with a 16-mm telephoto lens on a hunting rifle, a Raspberry Pi 3 to record the videos in the shooting field, specially designed targets with markers to help the marker detection via template matching, calculation of performance metrics and attributes (cf. \cref{subsec:implementation}), visualization, and compositing, as shown in \cref{fig:teaser}. 

\subsection{Design Goals \& Implementation} \label{subsec:implementation}
The composite visualizations were hypothesized to increase the coaches' and trainees' understanding of trainees' steadiness, accuracy, consistency, the recoil management, timing, pre-shot and post-shot behavior, and aiming trajectory. The design goals guiding the development included enabling comparative analysis, promoting self-reflection, and providing rapid terminal visual feedback to the trainee.
An agile development strategy was assumed, going through a few iterations of composite visualizations before deciding on the final ones for the user study.

The current implementation processes recorded videos offline using Python and OpenCV\footnote{https://opencv.org/} libraries. For target recognition and tracking, OpenCV's template matching functions were used on each frame. The necessary calculations were made in Python to obtain the attributes on time elapsed, distance of aimpoint from the target center, velocity and acceleration of the aimpoint, accuracy, and precision. As shown in \cref{fig:teaser}, the visualization and compositing were also done using OpenCV. Processing times were not formally benchmarked, but the design goal was to enable analysis within a few minutes post-recording.

\subsection{Proposed Composite Visualizations} \label{subsec:compositeVis}
Several iterative stages were passed through during design, which were mostly to understand and identify which of the calculated features are interesting according to the design goals described earlier and where and how to place them on the screen. The formative evaluation in these iterations was done by the authors by anticipating the needs of coaches and trainees. In the end, 
five visualizations were selected, as shown in \cref{fig:design_subfigs} and described briefly below.

\textbf{Vis~\#1 - Raw video footage} is the baseline or control group of the experiment. It is the simplest form of rapid terminal feedback (i.e., visual feedback that can be viewed right after the action), and both experts and non-experts are familiar with the visualization. 

\textbf{Vis~\#2 - Stabilized video with text overlay} removes recoil of the rifle \& presenting position and accuracy information as text.
That is, the video is stabilized to keep the target at the center at all times.

\textbf{Vis~\#3 - Time-series visualization of selected attributes} shows the difference between the aimpoint and the target center, its windowed mean, Euclidean distance, aimpoint acceleration, accuracy and precision in the last second graphs in a juxtaposed manner to understand steadiness, pre-shot, and post-shot behavior. 

\textbf{Vis~\#4 - A polar plot with the aim point and the target} shows the current aimpoint trajectory as well as the historical data (the last second in color and other historical locations in dashed gray) on a polar plot, simulating the projection of the aimpoint on target.

\textbf{Vis~\#5 - Combined composite visualization} juxtaposes three time-series visualizations from Vis~\#3 case (namely, accuracy in the last second, precision in the last second, and aim-point acceleration), polar plot (Vis~\#4), and raw video (Vis~\#1) with additional overlays of target highlight and aim point highlight. This last visualization uses two composite visualizations techniques within itself: juxtaposition and superimposition~\cite{javed2012exploring}.

All five composite visualizations relate to some or all of the aspects mentioned in the design goals, such as steadiness, accuracy, consistency, timing, pre-shot and post-shot behavior. Since this was intended as a prototype, no user interaction is available at this stage. Instead, the visualizations are provided as baked-in videos.

\section{Evaluation Study} \label{sec:evaluation}
A user study was conducted to evaluate the proposed system and identify new design directions for future composite visualization studies and products. Details of the study are provided below.

\subsection{Methodology} \label{subsec:methodology}
An experimental methodology was adopted to evaluate the developed alternative composite visualizations.

\textbf{Participants:} 10 participants were invited to a user study; 5 expert shooters (0F, 5M - who have actively practiced shooting for 6, 36, 40, 45, and 46 years), 5 novice participants (2F, 3M - with little to no experience in target shooting). 

\textbf{Raw Video Dataset:} Four 1-minute videos were used. Videos were recorded outdoors in a private property using the rifle setup shown in \cref{fig:teaser} by a professional shooter. 

\textbf{Procedure \& Tasks:} We used an embedded mixed-methods design~\cite{creswell2007chapter4}, in which a primarily quantitative data collection was complemented by embedded qualitative components, all conducted within a single session. The test procedure had three stages:

First, participants watched each of the above composite visualizations as videos, in a fixed order, following the numerical order. To avoid priming effects, all visualizations (except for Vis~\#1 and Vis~\#3, which either did not have video or visualizations) used a different raw video. Afterwards, each participant was asked to: rate how well they understood the shooter's aim in that video, determine how many shots there were, explain how they determined this number, and assess how well they understood the location of each shot on the target. 
The second stage was a pairwise comparison test, in which the participants were asked to indicate their preference when presented with two different visualizations (randomized order). 
Finally, a semi-structured interview was conducted in the end. 

Each user test took one hour on average, including watching the videos, responding to the questionnaire, and participating in a semi-structured interview. 
During the tests, visualizations were explained only when the participant did not understand them.

\begin{figure}
    \centering
    \includegraphics[width=0.99\linewidth]{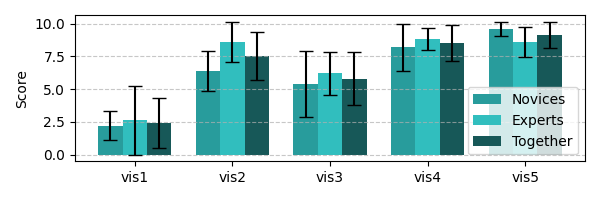}
    \caption{Mean and standard deviation for the participants' self-reported understanding how the shooter is aiming.}
    \label{fig:understandingResult}
\end{figure}

\textbf{Data Analysis Methods:} For the rating task of understanding how the shooter is aiming, two non-parametric tests were used to obtain more robust results considering the small number of participants: Mann-Whitney U and Wilcoxon signed-rank tests. For pairwise comparisons results, three different analysis methods were used. First, the coefficient of consistence, $\zeta$, was computed to identify whether the groups were consistent within themselves considering circular triads~\cite{banterle2009psychophysical}. 
Test of equality, $D$, was used to determine whether the score differences were statistically significantly different overall~\cite{david1963method}. 
Multiple comparisons test was then used to calculate $R$, which acts as a threshold value for difference between pairs to determine statistical significance between them~\cite{banterle2009psychophysical, selmanovic2013generating}.

\subsection{Quantitative Results} 
Participants were asked \textit{``How well do you understand how the shooter is aiming in the video?''} to determine the effect of the visualization on their understanding. \cref{fig:understandingResult} shows the results for novices, experts, and both groups combined. A pairwise Mann-Whitney U test shows that the differences between novices and experts are not statistically significant. Wilcoxon signed-rank tests showed that only Vis~\#1 was found statistically worse than Vis~\#2, \#4, and \#5. 

Participants' preferences in the pairwise comparisons test are reported in \cref{tab:pwcResults}, where the results for novices and experts are differentiated. Summarized in \cref{tab:significanceTable}, statistical test results show that there are no circular triads ($\zeta=1$), and the groupings show statistical significance according to the multiple comparisons test. Both novices and experts found the combined composite visualization (Vis~\#5) the most preferred, although this finding is not significant within the expert group. 

\begin{table}[tb]
  \caption{Results of the pairwise comparison results. The value on each cell indicates that the visualization stimulus on that row was preferred by that many participants against the visualization on that column. The color and bold indicate results for ``Novice$|$\expv{Expert}''.}
  \label{tab:pwcResults}
  \scriptsize%
	\centering%
  \begin{tabu}{%
    l|%
    *{5}{c}|%
    c%
    }
  \toprule
       &   Vis1 & Vis2 & Vis3 & Vis4 & Vis5 & \multirow{2}{*}{Total} \\ 
       &    Raw & Text & Plots& Polar& Comb. &  \\ 
  \midrule
  Vis1 - Raw   &  - &  0$|$\expv{0} & 0$|$\expv{1} & 0$|$\expv{0} & 0$|$\expv{0} 
       & 0{~}$|$\expv{{~}1} \\
  Vis2 - Text  &  5$|$\expv{5} &  - & 3$|$\expv{3} & 2$|$\expv{2} & 0$|$\expv{2} 
       & 10$|$\expv{12} \\
  Vis3 - Plots &  5$|$\expv{4} &  2$|$\expv{2} & - & 2$|$\expv{1} & 0$|$\expv{1} 
       &  9{~}$|$\expv{{~}8} \\
  Vis4 - Polar &  5$|$\expv{5} &  3$|$\expv{3} & 3$|$\expv{4} & - & 0$|$\expv{0} 
       & 11$|$\expv{12} \\
  Vis5 - Comb. &  5$|$\expv{5} &  5$|$\expv{3} & 5$|$\expv{4} & 5$|$\expv{5} & - 
       & 20$|$\expv{17} \\
  \bottomrule
  \end{tabu}%
\end{table}


\begin{table}[b]
  \caption{Ranking and statistical significance results of pairwise comparison stage of the user study. The consistency score $\zeta=1$ for all cases. All $D$ values show significance; $D\geq \chi^2(4,0.05)\approx9.448$ for $D_{Novice, Expert}$ and $D_{Combined}\geq \chi^2(9,0.05)=16.919$. Groupings show significance according to the multiple comparisons test.}
  \label{tab:significanceTable}
  \scriptsize
	\centering
  \begin{tikzpicture}[overlay, remember picture]
  \node (start) at (0,0) {}; 
  \end{tikzpicture}
  
  \begin{tabu}{l c c|*{5}{c}}
  \toprule
   \multirow{2}{*}{} & \multirow{2}{*}{$\zeta$} & \multirow{2}{*}{$D$} & \multicolumn{5}{c}{Rankings} \\
    & & & $1^{st}$ & $2^{nd}$ & $3^{rd}$ & $4^{th}$ & $5^{th}$ \\ 
   \midrule
   Novice   & 1 & 32.32 & 
       \tikzmarknode{novvis1}{Vis5} & 
       \tikzmarknode{novvis2}{Vis4} & 
       \tikzmarknode{novvis3}{Vis2} & 
       \tikzmarknode{novvis4}{Vis3} & 
       \tikzmarknode{novvis5}{Vis1} \\
   Expert   & 1 & 22.72 & 
       \tikzmarknode{expvis1}{Vis5} & 
       \tikzmarknode{expvis2}{Vis2} & 
       \tikzmarknode{expvis3}{Vis4} & 
       \tikzmarknode{expvis4}{Vis3} & 
       \tikzmarknode{expvis5}{Vis1} \\
   \midrule
   All & 1 & 53.76 & 
       \tikzmarknode{comvis1}{Vis5} & 
       \tikzmarknode{comvis2}{Vis4} & 
       \tikzmarknode{comvis3}{Vis2} & 
       \tikzmarknode{comvis4}{Vis3} & 
       \tikzmarknode{comvis5}{Vis1} \\
   \bottomrule
   \end{tabu}

\begin{tikzpicture}[remember picture,overlay]
  \draw[deepblue, thick, dashed, rounded corners=3pt]
    ([yshift=1pt]novvis1.south west) rectangle ([yshift=-1pt]novvis2.north east);
  \draw[skyblue, thick, rounded corners=3pt]
    ([yshift=1pt]novvis2.south west) rectangle ([yshift=-1pt]novvis4.north east);
  \draw[iceblue, very thick, dotted, rounded corners=3pt]
    ([yshift=1pt]novvis4.south west) rectangle ([yshift=-1pt]novvis5.north east);
  \draw[skyblue, thick, rounded corners=3pt]
    ([yshift=1pt]expvis1.south west) rectangle ([yshift=-1pt]expvis4.north east);
  \draw[iceblue, very thick, dotted, rounded corners=3pt]
    ([yshift=1pt]expvis4.south west) rectangle ([yshift=-1pt]expvis5.north east);
  \draw[skyblue, thick, rounded corners=3pt]
    ([yshift=0pt]comvis2.south west) rectangle ([yshift=0pt]comvis4.north east);
\end{tikzpicture}
\end{table}

\subsection{Qualitative Findings} 
The collected qualitative data provide insight into how well the visualizations assist participants in understanding and analyzing the shooting process. Some comments were common in both groups, while novices and experts differed in others. 

Novices' comments focused mostly on the aesthetics and visual representations, instead of functionality. 
In a few cases, they mentioned that they disliked the colors, disliked the graphical design, and wanted improvements on visuals. There were some conflicting comments for Vis~\#3; some participants liked it very much and some others disliked strongly.  Their comments also showed that recoil was an important tool in understanding how many shots there were, as well as aimpoint focusing and slowing down onto a single point. Among others, Vis~\#4 and \#5 were found easier to follow.

Expert shooters praised the recordings and composite visualizations mentioning this can be a great tool for marksmanship training (cf. \cref{subsec:coaching}). They liked the idea of having data shown, but could not read or understand the text data on the side for Vis~\#2. They found the acceleration graph very useful, mentioning that it was also easy to understand. Regarding polar plot (Vis~\#3), experts wanted the ability to filter the lines (i.e., old or irrelevant data) to look at a specific one. Experts have mentioned in a few instances (i.e., Vis~\#3 and \#4) that these graphs could be used together with the video recordings, essentially foreshadowing Vis~\#5 (which they have not seen at that point). For the combined composite visualization (Vis~\#5), notable comments mentioned that ``it [gets] easier to understand the graphs the more [one] uses them''. They also indicated that ``there is too much going on in the video at the same time'', indicating the sensory and cognitive overload, which necessitates the users of this tool to learn the tool beforehand.

Participants across expertise levels found the composite view (Vis~\#5) more informative and supportive than isolated visualizations (Vis~\#2–Vis~\#4). Regardless of the skill level, participants requested an indication of where the shots hit the target, which can be implemented later on. Other requests were an indication for distance from the target and an information box where details about the weapon, shooter, other details are provided. Both groups mentioned that Vis~\#2 and Vis~\#3 were not easy to follow. Both groups liked the use of colors to highlight important things (e.g., aimpoint is on target, and the tail of aimpoint history) and how recorded video and visualizations are shown together. All these comments point out the strengths and weaknesses of the developed visualizations, which inform design choices as discussed more in the next section.

\section{Discussion} \label{sec:discussion}




\subsection{Coaching Perspective} \label{subsec:coaching}

Expert participants of the user study considered the combined composite visualizations with first-person video recording very beneficial, as it would help to observe the shooter's actions and correct faults happening during shooting. In particular, they indicated that a video would enable them to see how the aim is affected by breathing, see the user's aiming prior to firing, and compare the video to the hit marks. 
Moreover, they thought this tool will be useful to teach shooting instructors, too. These comments show that similar video-based composite visualizations can be used in other sports training applications to enhance effectiveness of the training.

The experts also mentioned that it would have been nice to see the video of the shooter's body and face, since they thought seeing body posture and breathing is also interesting, which is in line with the prior literature describing how the target shooting teaching in practice is done~\cite{knight2007perfect, johnson2008crucial, brown2018stance}. Such multiple recordings can easily be made and will benefit composite visualizations for other sports. 

The aimpoint acceleration graph, which was the visual output of attribute extraction from video and subsequent computations, was found extremely useful by the expert shooters. Similar analyses are applicable on extracted features to identify flaws and suggest better training regimes~\cite{hoferlin2010video}.

\subsection{Design Implications} 
The qualitative comments from participants shine light into design implications for future visualization design studies. 

The diverging comments from novices and experts show that the requirements are different for learners and instructors. Novices want customization options for colors and visual aesthetics. Experts mentioned that there is a need to filter data. When designing a visual interface, both requirements and expectations of the user need to be taken into consideration, depending on the skill level~\cite{wassink2009applying, elias2011exploration}. This finding indicates a need for having two versions of such visual analytics tools, one aimed at trainees and one aimed at instructors. In time, trainees can also be given access to the advanced settings.

Despite other differences, both groups wanted more information about where the bullets hit, the weapon used, distance to the target, and the details of the shooter. Moreover, overlapping comments from both groups indicated the final visual stimuli should not be overwhelming with too many graphs or lines. These bring forward how crucial an iterative design cycle is~\cite{sedlmair2012design}, and not with only usability tests but also with semi-structured or unstructured interviews.

The realistic visual representation of recoil was a very important cue for participants in both groups. This shows how important the real video recording is in understanding the performance of the shooter. The importance of video-recordings have been shown in other sports, too~\cite{reichert2021stabilization, hoferlin2010video}. In future visual analytics frameworks for sports training applications, the first-person (or third-person) video recordings should be the basis for motion analysis and visualization will bring additional understanding into the training procedure. The insights gained in this particular project can also be applied to archery (with hit confirmation, aiming prior to shooting, offset between the target point and actual shot location), baseball (swing trajectory, timing, aiming bat at correct location), or other sports.

\section{Limitations \& Future Work} \label{sec:limitatinos}

This evaluation study involved a limited number of participants albeit with different skill levels. While this does not invalidate the findings, a larger study would improve the generalizability of the conclusions. We plan to continue studying the composite visualizations for sports technologies. 

There are two technical limitations that can be resolved easily. First, the current system records video on a Raspberry Pi while computations are performed on a separate laptop; this can be simplified by using a laptop directly in the field. Second, video drift occurred because the camera was not rigidly attached to the rifle, which can be fixed by proper mounting via drilling.

Mounting an additional camera on the rifle may have affected its load balance and the shooting experience. This should be further analyzed to optimize camera placement.

In its methodology, the study used pre-rendered videos rather than an interactive visualization, which would likely offer greater training benefits. Still, the baked-in videos can provide crucial rapid terminal visual feedback with minimal processing time. The order of visualizations was not randomized, which may introduce ordering bias or learning effects. However, no such effects were observed in the results or in the qualitative feedback.

Finally, in this prototype, visual aesthetics, color choices, and refinements were not prioritized. 
Future studies can focus on improving the graphical design to enhance user experience.

\section{Conclusion} \label{sec:conclusion}
Effective training in precision sports requires timely, interpretable feedback that connects athletes' actions to performance outcomes. In this paper, we provide a design study to develop composite visualizations for marksmanship training application which provides insight into the design implications for future sports training visualization applications. We further discuss the coaching perspective and how visualizations can benefit other instances of sports training. 
Our study confirms that combined composite visualizations showing video recordings, rather than visualizations detached from the source video, best support rapid visual feedback in marksmanship training. This finding is supported by both the self-evaluated understanding of how the shooter is aiming and by the pairwise-comparisons test for their preference. The developed system, therefore, is a fruitful attempt in fulfilling the unmet need of having composite visualizations for sports training applications. Ultimately, this work lays the groundwork for integrating lightweight, video-driven analytics into everyday sports training, bridging the gap between expert intuition and actionable visual feedback.

\acknowledgments{
This work was partially supported by the Knowledge Foundation, Sweden, with grant number 2019-0251.}

\bibliographystyle{abbrv-doi}

\bibliography{main}
\end{document}